# Hierarchical Distribution Matcher Design for Probabilistic Constellation Shaping Based on a Novel Semi-Analytical Optimization Approach

Pantea Nadimi Goki, *Member, IEEE*, Luca Potì, *Senior Member, IEEE*

***Abstract*—A novel design procedure for practical hierarchical distribution matchers (HiDMs) in probabilistically shaped constellation systems is presented. The proposed approach enables the determination of optimal parameters for any target distribution matcher rate $R_{DM}$. Specifically, lower bounds on energy loss, rate loss, and memory requirements are analytically estimated for HiDM architectures approximating the Maxwell–Boltzmann (MB) distribution. A semi-analytical optimization framework is employed to jointly optimize rate and energy loss, allowing the selection of the number of hierarchical layers, memory size, and block length required to optimize channel capacity. The accuracy of the proposed model is validated through probabilistic amplitude shaping of 16-QAM (PAS-16QAM), showing good agreement between analytical predictions and simulated results. The proposed analytical tool facilitates the design of HiDM structures that are compatible with practical hardware and implementation constraints, such as those imposed by state-of-the-art application-specific integrated circuits (ASICs) and field-programmable gate arrays (FPGAs). Furthermore, the performance of the optimized HiDM structure, incorporating layer selection based on lower-bound energy loss, is evaluated over the AWGN channel in terms of normalized generalized mutual information (NGMI) as a function of the optical signal-to-noise ratio (OSNR). At a net data rate of 200 Gb/s with 25% forward error correction (FEC) overhead, the proposed scheme achieves a shaping gain improvement of 2.8% compared to previously reported solutions.**

## I. INTRODUCTION

Probabilistic amplitude shaping (PAS) is a groundbreaking technique that enables the development of energy-efficient, high-capacity optical communication systems. It offers a promising solution to the considerable challenges of designing high-performance, energy-optimized systems for today's globally extended communication networks.

*(Corresponding author: Pantea Nadimi Goki).*

Pantea Nadimi Goki is with the Telecommunications, Computer Engineering, and Photonics (TeCIP) Institute, Scuola Superiore Sant'Anna, 56124 Pisa, Italy, and with the National Laboratory of Photonic Networks, CNIT, 56124 Pisa, Italy (e-mail: pantea.nadimigoki@santannapisa.it).

Luca Poti is with the Faculty of Technological Sciences and Innovation, Universitas Mercatorum, Piazza Mattei10, 00186 Roma, Italy, and with the National Laboratory of Photonic Networks, CNIT,56124 Pisa, Italy (e-mail: *luca.poti@cnit.it*)

PAS supports the pursuit of optimal coherent systems with high information rates, low cost, strong noise tolerance, and the ability to extend transmission distances while maintaining high spectral efficiency. Its importance is underscored by the limitations of uniform quadrature amplitude modulation (U-QAM) formats, such as 16-QAM, which face significant challenges in achieving optimal performance at high data rates. Optimizing the mutual information (MI) between the channel input and output enables a communication system to operate closer to its capacity, allowing for higher data rates. In information theory, this is known as constellation shaping (CS) [1], which seeks to improve energy efficiency over conventional coherent transmission [2]. Probabilistic amplitude shaping (PAS) is a prominent CS method that optimizes modulation to approach the Shannon limit. The shaping gap, the MI loss of uniform signaling, limits achievable rates. PAS provides a shaping gain that reduces this gap and increases capacity. It approaches the additive white Gaussian noise (AWGN) channel capacity under an average power constraint by adopting a Maxwell-Boltzmann (MB) distribution, the discrete counterpart of the Gaussian distribution for constellation inputs [3]. PAS lowers the required transmission energy for a given rate or supports higher rates at a given signal-to-noise ratio (SNR). By assigning higher probabilities to low-energy symbols and lower probabilities to high-energy symbols [4], PAS maximizes entropy for a given average energy, thereby reducing the average symbol energy and improving spectral efficiency.

Extensive research has been conducted on implementing probabilistic amplitude shaping in coherent systems (PAS-QAM), which can be classified into two main approaches: Indirect PAS (I-PAS) and Direct PAS (D-PAS). I-PAS, also known as *sphere shaping*, targets a specific rate by confining the $n$-dimensional signal structure within a spherical boundary [5,6] to maximize energy efficiency for that rate. Prominent I-PAS architectures include enumerative sphere shaping (ESS) and shell mapping (SM), both using trellis-based mapping with different indexing algorithms [7-9]. D-PAS, also known as *distribution matching*, employs a distribution matcher (DM) to target a predefined probability distribution, typically Maxwell–Boltzmann, over low-dimensional signal structures [6]. It aims to achieve maximum capacity with minimal energy for the specified rate. D-PAS is generally more practical and less complex, whereas I-PAS can achieve higher efficiency at the cost of increased computational complexity and memory requirements. Both methods adapt the information rate to the channel SNR, reducing the gap to the Shannon limit, though their achievable rates differ.

The main challenge in D-PAS is designing an optimal distribution matcher (DM) that maps $k$ uniform input bits to $N$-shaped output amplitudes, achieving a target distribution with entropy $H$ (in bits per amplitude), matching the output amplitudes. The DM rate is defined as $R_{DM} = k/N$ (in bits per amplitude), and the rate loss is $R_{loss} = H - R_{DM}$ (in bits per amplitude) reflecting the difference between the output entropy ($H$) and the actual ($R_{DM}$) entropy of the output distribution. Optimal DM minimizes $R_{loss}$. Studies show that I-PAS generally yields lower $R_{loss}$ than D-PAS [6]. In D-PAS, the DM introduces correlation between shaped amplitudes, reducing the achievable information rate compared to independent and identically distributed (i.i.d.) amplitudes with the same probability distribution [10]. Various DM architectures have been proposed [11,12], including: constant composition distribution matcher (CCDM) an invertible fixed-length DM using arithmetic coding [12]; multiset partition DM (MPDM) which employs multiple compositions and shorter DMs resulting in lower rate loss than CCDM [13]; arithmetic coding (AC) a fixed-to-fixed invertible DM with an energy-based approach [14]; two-parallel DM (2PaDM) which employs two distribution matchers (DMs) operating in parallel, achieves a substantial memory reduction and is compatible with the standardized Flexible Optical Transport Network [15]. A recent and impactful method is hierarchical DM (HiDM) [16], which combines multiple small look-up tables (LUTs) in a hierarchical structure using index mapping. HiDM is operation-free, requiring only LUT access [10], and offers high flexibility with respect to hardware-specific constraints. Properly designed HiDM significantly reduces $R_{loss}$, achieving lower inverse-DM error rates than CCDM [17]. HiDM design is constrained by LUT input/output bits $N_b$, the number of layers $L$, and the block length $N$. While $R_{loss}$ is the primary PAS performance metric, it does not reflect constellation energy. When targeting an MB distribution, the energy loss $E_{loss}$ can be used instead [10].

In this work, we develop a semi-analytical model of the constellation energy $E$ for HiDM, targeting an MB distribution, and evaluating performance using $R_{loss}$ and $E_{loss}$. We show that for a given $N_b$, increasing $L$ or $N$ beyond certain thresholds yields negligible gains, and we derive an equation for the mean per-amplitude energy (1D Symbol) for a given $L$ ($L > 4$). This enables $E_{loss}$ computation for any $L$ and $N$, starting from a 4-layer optimized design determined via exhaustive search. We demonstrate that HiDM $R_{loss}$ remains above its lower bound and does not vanish with currently available hardware; equivalently, mean energy remains above its lower bound. This powerful analytical tool is used for system design under practical constraints; finally, we evaluate transmission performance in terms of NGMI on an AWGN channel. This paper is organized as follows. Section II provides a brief overview of the key parameters and structure of hierarchical distribution matchers (HiDMs), together with a description of the proposed semi-analytical method for optimizing HiDM parameters through rate and energy loss reduction and decoding memory estimation. Section III presents the performance evaluation and comparison with existing results in the literature. Finally, Section IV concludes the paper.

## II. Model and Design

Implementing PAS in a network system requires three key properties: (i) practicality, (ii) scalability, and (iii) efficiency. A carefully designed HiDM can satisfy all three. In [10], a LUT-based HiDM architecture was introduced, featuring multiple layers $L$, each composed of small LUTs storing disjoint amplitude sequences ordered by ascending energy. In [18], we proposed a design where parameters are chosen to minimize $R_{loss}$ for a given $R_{DM}$. Previous works [10,16] reported that increasing $N$ reduces $R_{loss}$, equivalently lowering the mean symbol energy. Here, we extend our model, including design constraints, and we provide analytical expressions for $E_{loss}$ and $R_{loss}$ as an effective tool for system optimization. In the following, we adopt the notation, DM architecture, and algorithms from [10].

### *A. Hierarchical distribution matcher and LUT design constraints*

A HiDM structure is defined and implemented by a set of characterization vectors. Optimizing these vectors, which represent the entire $L$-layer HiDM structure, is critical to achieving an optimal distribution matcher for a one-dimensional target rate. These vectors are defined as follows.

-*Input bits*: $\boldsymbol{k} = (k_1, k_2, \dots, k_L)$, this vector represents the number of input bits to each LUT in each layer of the structure, where the subscript indicates the $l^{th}$ layer of the structure. The total number of bits mapped by the $L$-layer structure is given by [10]

$$k = \sum_{l=1}^{L} k_l T_l \qquad (1)$$

where, $T_l$ is the number of times that each layer is used ($T_l = \prod_{h=l+1}^{L} N_h$ with $T_L = 1$), due to the full parallelization of the HiDM structure performance.

-*Output amplitudes*: $\boldsymbol{N} = (N_1, N_2, \dots, N_L)$, this vector indicates the block length of LUTs in each layer. Each layer consists of small LUTs of the same size, i.e., the same number of sequences ($2^{k_l}$) and the same number of amplitudes per sequence ($N_l$) per LUT. There is no shared sequence between LUTs, making each LUT unique. The total output-shaped amplitudes from the $L$-layer HiDM structure, i.e., the DM word, is given by

$$N = \prod_{l=1}^{L} N_l \qquad (2)$$

-*Alphabet size*: $\boldsymbol{M} = (M_1, M_2, \dots, M_L)$, this vector represents the alphabet size of each layer, where $M_1$ indicates the $4M_1^2$-QAM constellation (e.g., when considering both the sign and the in-phase and quadrature components) alphabet size in the first layer, and $M_{l+1}$ is equal to the number of LUTs in layer $l$, except for the last layer, which only has one LUT, so that the number of LUTs can be expressed by the vector $\boldsymbol{U} = (M_2, M_3, \dots, M_{L-1}, 1)$.

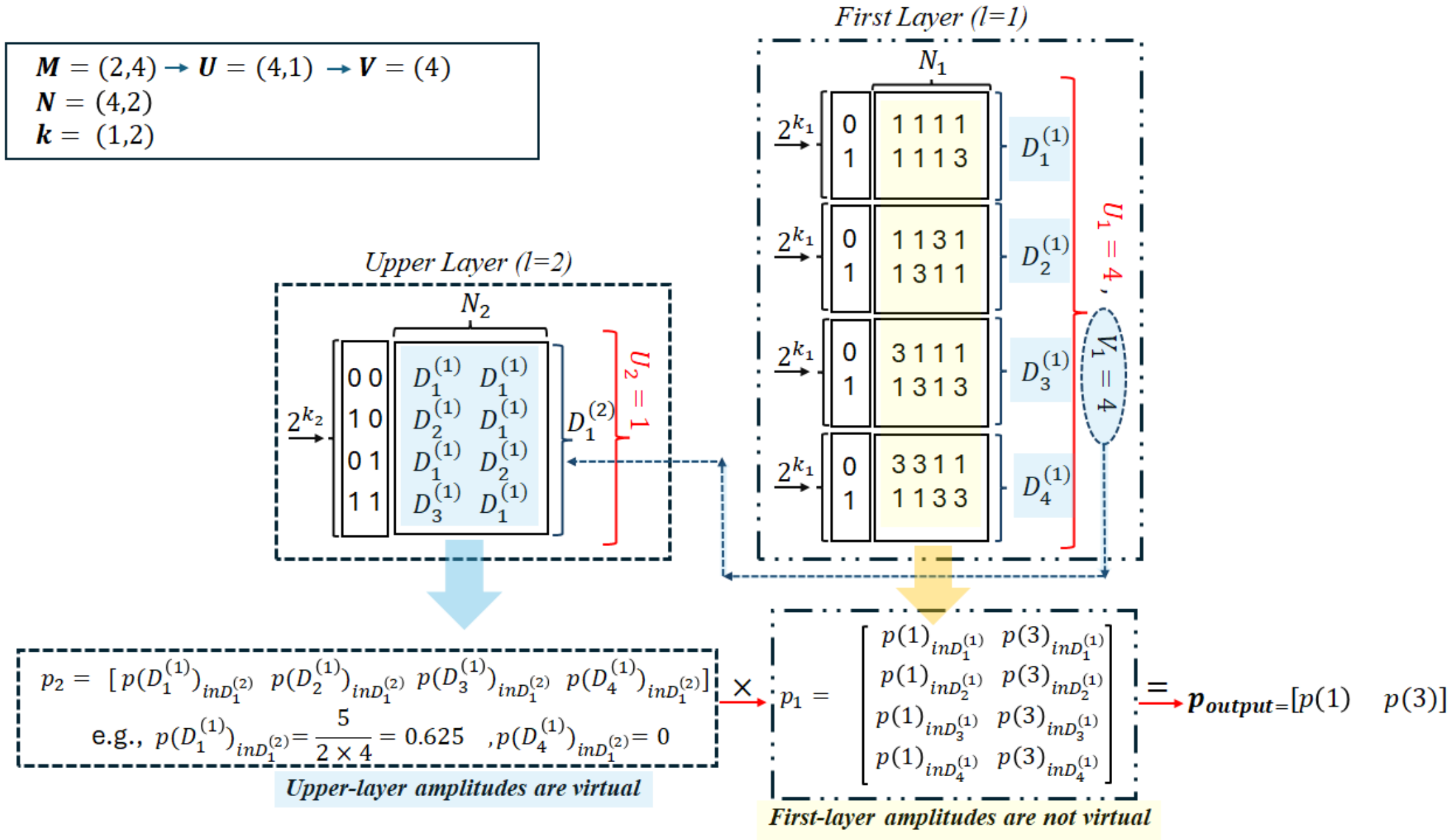

**Fig.1.** A two-layer HiDM structure, LUT-based DM with the following characterization vectors $\boldsymbol{M}$= *(2,4)*, $\boldsymbol{N}$= *(4,2)*, $\boldsymbol{k}$= *(1,2)*. Below, it shows the map of the recursive calculation of output amplitudes $p$ probabilities. $V_l$: virtual amplitude, $U_l$: number of LUTs (in the $l^{th}$ layer). A detailed explanation of this example is provided in Section *B*.

Moreover, the LUTs of each layer will be utilized as a virtual amplitude by the upper layer, identified by the vector $\boldsymbol{V} = (M_2\,, M_3, \ldots, M_L)$, which starts from the second layer since the first layer includes QAM amplitudes.

For each upper layer ($l \geq 2$), a $2^{k_l}U_l$ sequences of length $N_l$from the set of $D^{(l-1)} = \left\{D_1^{(l-1)}, D_2^{(l-1)}, \ldots, D_{U_l}^{(l-1)}\right\}$ is generated, sorted by energy, and each $2^{k_l}$ sequence is placed in each $D_y^{(l)}$, where $y$ indicates the DM number in the given layer. While the number of LUTs in each layer can be chosen based on system requirements, the last layer $L$ always contains a single LUT, which will be utilized only once, $T_L = 1$ . A simple example of a two-layer HiDM structure is shown in Fig. 1, where $k = k_1 T_1 + k_2 = 4$ input bits and a rate of $R_{DM} = 0.5$ so that the output amplitudes will be 8. Here, the upper layer ($l = L = 2$) is used first and one time ($T_2 = 1$), while the first layer ($l = 1$) is used twice ($T_1 = 2$).

Different combinations of the characterization vectors, having the same $R_{DM}$ and $N$, could result in varying $H$ thus giving different $R_{loss}$. Finding optimized characterization vectors to minimize $R_{loss}$ and $E_{loss}$ at a given rate is a major challenge in HiDM design. An exhaustive search can identify optimal structure parameters but demands significant hardware resources. Such a search is feasible only for a limited number of layers and for a small set of $\boldsymbol{M}$ and $\boldsymbol{N}$. However, as $L$ and $\boldsymbol{M}$ grow, exhaustive search becomes impractical. Thus, efficient optimization techniques are required to determine the optimal parameters.

To provide a simple illustrative example, we consider the generation of the optimal 4-layer HiDM structure for implementing 1D PAS with two amplitude levels (corresponding to 16QAM modulation format), targeting $R_{DM} = 0.5$ (bits/amp). Using the architecture and algorithm described in [10], we performed an exhaustive search over all possible 4-layer HiDM configurations without constraints on $\boldsymbol{N}$, but restricting $N_b \leq 12$ and $M_{l>2} \in \{16,32,64\}$ since lower values provide high $R_{loss}$ and higher values require too large computational memory for exhaustive search. In this case, the structure is defined by $\boldsymbol{M} = (2, M_2, M_3, M_4)$, $\boldsymbol{N} = (N_1, N_2, N_3, N_4)$, and $\boldsymbol{k} = (k_1, k_2, k_3, k_4)$, resulting in 27 possible combinations for $\boldsymbol{M}$ and any acceptable values for $N_i$ and $k_i$.

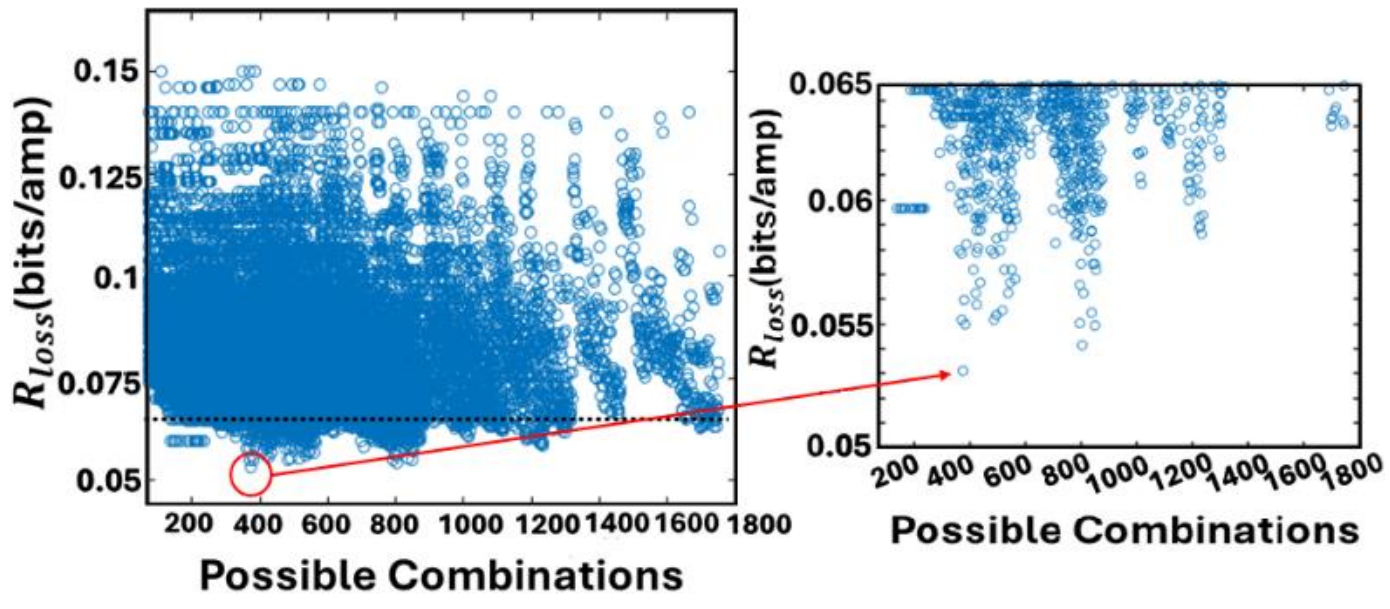

**Fig. 2.** An exhaustive search for PAS-16QAM with $R_{DM} = 0.5$ evaluated 20,417 structures in terms of $R_{loss}$. The zoomed plot shows the area below the dashed line that corresponds to the results obtained with $\boldsymbol{M} = (2, \bar{M}, \bar{M}, \bar{M})$, where $\bar{M} \in \{16,32,64\}$, and $\boldsymbol{N} = (N_1, n, n, n)$. The red arrow highlights the lowest achievable $R_{loss}$ with **M** = (2,64,64,64), $\boldsymbol{N}$= (11,2,2,2), and **k** =(2,3,4,8).

Using these parameters, in Fig.2, the $R_{loss}$ values for all possible $N_i$ and $k_i$ combinations for each set of $\boldsymbol{M}$ are reported. The total number of HiDM structure configurations is 20,417. The area below the dashed line corresponds to the results obtained with $\boldsymbol{M} = (2, \bar{M}, \bar{M}, \bar{M})$, where $\bar{M} \in \{16,32,64\}$ and $\boldsymbol{N} = (N_1, n, n, n)$, where $N_1 > n$. Among them, a single combination achieves the best performance, highlighted in the zoomed inset, with parameters: $\boldsymbol{M} = (2, 64,64,64)$, $\boldsymbol{N} = (11,2,2,2)$, $\boldsymbol{k} = (2,3,4,8)$; yielding a $R_{loss} = 0.053$ (bits/amp) with $N$=88. As the number of layers increases, such an exhaustive approach becomes computationally unfeasible.

Here, we propose a method to determine optimal parameters for any specific $R_{DM}$. Our methods can be easily adapted to practical cases where design tools and hardware limitations further reduce parameters ranges. While commercial tools for field-programmable gate array (FPGA) synthesis can handle only a limited number of inputs per single physical look-up table, they are designed to decompose complex logic functions to fit the device's architecture. In modern application-specific integrated circuits (ASICs), however, hardware limitations are related to the physical area of the chip, with on-chip memory being a major determinant of the final die size and manufacturing cost. In HiDM implementations, the maximum number of input/output bits for each LUT is $N_b = \max\{N_i \log_2 M_i, k_i + \log_2 U_i\}$ where the terms represent DM output and invDM input bits, respectively, for the $i^{\text{th}}$ layer. Commercial tools for LUT synthesis today work fine with $N_b \leq 12$. At the same time, a reasonable memory limit is in the order of hundreds of Mbit. In our former example, the maximum number of bits was $N_b = 12$ and the memory $mem = 467968$ bits. This approach imposes the following constraints on the characterization vectors to design the HiDM structure.

(i) $\boldsymbol{M} = (M_1, \bar{M}, \ldots, \bar{M})$, this constraint enforces equal alphabet size across all layers except the first. Increasing the alphabet size in an upper layer results in a corresponding increase in the number of sequences in the lower layer.

(ii) $\boldsymbol{N} = (N_1, n, \ldots, n)$ where $N_1 > 2$. In this scheme, most amplitudes are allocated to the first layer, while the remaining amplitudes are evenly distributed among the other layers. Assigning a long block length ($N_1$) to the first layer and a large number of virtual alphabets to subsequent layers enables mapping bits onto long sequences of low-energy amplitudes. Sequences with a high probability of smaller amplitudes better approximate the target Maxwell-Boltzmann distribution.

(iii) $M_l^{N_l} \geq U_l 2^{k_l}$ ensures enough sequences for filling all DMs.

(iv) $N_l \log_2 M_l \geq k_l$ ensures the number of output bits is larger than the number of input bits of each DM in each layer.

The two latest constraints (iii) and (iv) guarantee DM invertibility. The proposed method assumes a specific value for the maximum number of bits $N_b$, defines the maximum $\bar{M}_{max} = 2^{\left\lceil \frac{N_b}{n} \right\rceil}$ and performs an exhaustive search to find one characterization vector (i.e., $\boldsymbol{k}$), starting with $N_1 = N_b$. The search varies $N_1$ from $N_b$ down to 1, stopping when $R_{loss}$ at $N_1 = N_b - (c-1)$ exceeds that at $N_1 = N_b - c$, with $c \in \mathbb{N}$. The value $N_1 = N_b - c$ is used to determine the optimal $\boldsymbol{k}$.

For example, for $N_b = 12$, $\boldsymbol{M} = (2,64,64,64)$, $\boldsymbol{N} = (N_1, 2,2,2)$, an exhaustive search over $\boldsymbol{k}$ yielded the optimal structure: $\boldsymbol{k} = (2,3,4,8), N_1 = 11$, with $R_{loss} = 0.053$ (bits/amp) at $N$=88. This result matches the exhaustive search for $\boldsymbol{M}$, $\boldsymbol{N}$, and $\boldsymbol{k}$, while significantly reducing the parameters research space down to 294.

For the remainder of this paper, we consider PAS-16QAM with parameters: $M_1 = 2$ , $n = 2$, $\bar{M} = 2^{\left\lceil \frac{N_b}{2} \right\rceil}$ , and $N_1 = (N_b - 1)$. While not claiming generality, these values represent the optimal choice based on an exhaustive parameter search up to 14 layers. This configuration effectively reduces $R_{loss}$ and, consequently, $E_{loss}$.

### *B. Probability and Energy Loss of 16QAM-HiDM*

The DM output amplitudes $\boldsymbol{x}_1 \in \mathcal{A} = \{1,3\}$ have probabilities $p(1) = p$ and $p(3) = 1 - p$ , where $p$ is calculated recursively as in [10]. The recursion starts from the outermost layer $L$ with $p_L(x)$ and proceeds inward to the first layer, yielding $p$, as illustrated in Fig. 1 . As described in Section *A*, for $l \geq 2$, $x$ denotes the specific DM in layer $l-1$, defined as virtual amplitude ($V_l$) for the layer $l$, and is placed in the DM at layer $l$, denoted $D_y^{(l)}$. The conditional probability that $x$ is generated at layer $l$, given that $y$ was chosen at layer $l$+1 is given by [10]:

$$p_{l|l+1}(x|y) = \frac{\# of\ occurance\ of\ x\ in\ D_y^{(l)}}{N_l 2^{k_l}} \tag{3}$$

The probability of $x$ after passing through layers *L*, *L*-1,...,1 is given by:

$$p_l(x) = \sum_{y \in \{1,2,\ldots,M_{l+1}\}} p_{l|l+1}(x|y) p_{l+1}(y) \tag{4}$$

Ultimately, the output probability of $\boldsymbol{x}_1$ is obtained recursively, from the last layer to the first, using the law of total probability [10]. Knowing the probabilities of the output amplitudes enables calculation of the entropy and the mean energy per 1D symbol of the structure's output. The mean energy per amplitude of PAS-16QAM is given by:

$$E_{DM} = \sum_{x_1 = 1,3} P(x_1) x_1^2 = (9 - 8p) \tag{5}$$

While $R_{loss}$ is an intrinsic property of HiDM; the shaping gap is computed relative to an MB distribution with entropy equal to $R_{DM}$ , as an ideal CS with optimal shaping rate $R_{DM} = H$ [19]. The energy gap [20], known as energy loss (in dB), is defined as the ratio between the HiDM mean energy per amplitude and that of the MB distribution [10]:

$$E_{loss} = 10 \log_{10} \left( \frac{E_{DM}}{E_{MB}} \right) \tag{6}$$

### *C. $E_{loss}$ and $R_{loss}$ Semi-analytical expression*

We outline a simple procedure to estimate the asymptotic limit of the mean energy per amplitude and the rate loss in a practical HiDM design for a 16QAM probabilistically constellation-shaped system targeting the MB distribution. By empirically examining various HiDM structures, we observed that the mean energy per amplitude, and equivalently $R_{loss}$, decreases as the number of layers $L$ increases for a fixed $R_{DM}$.

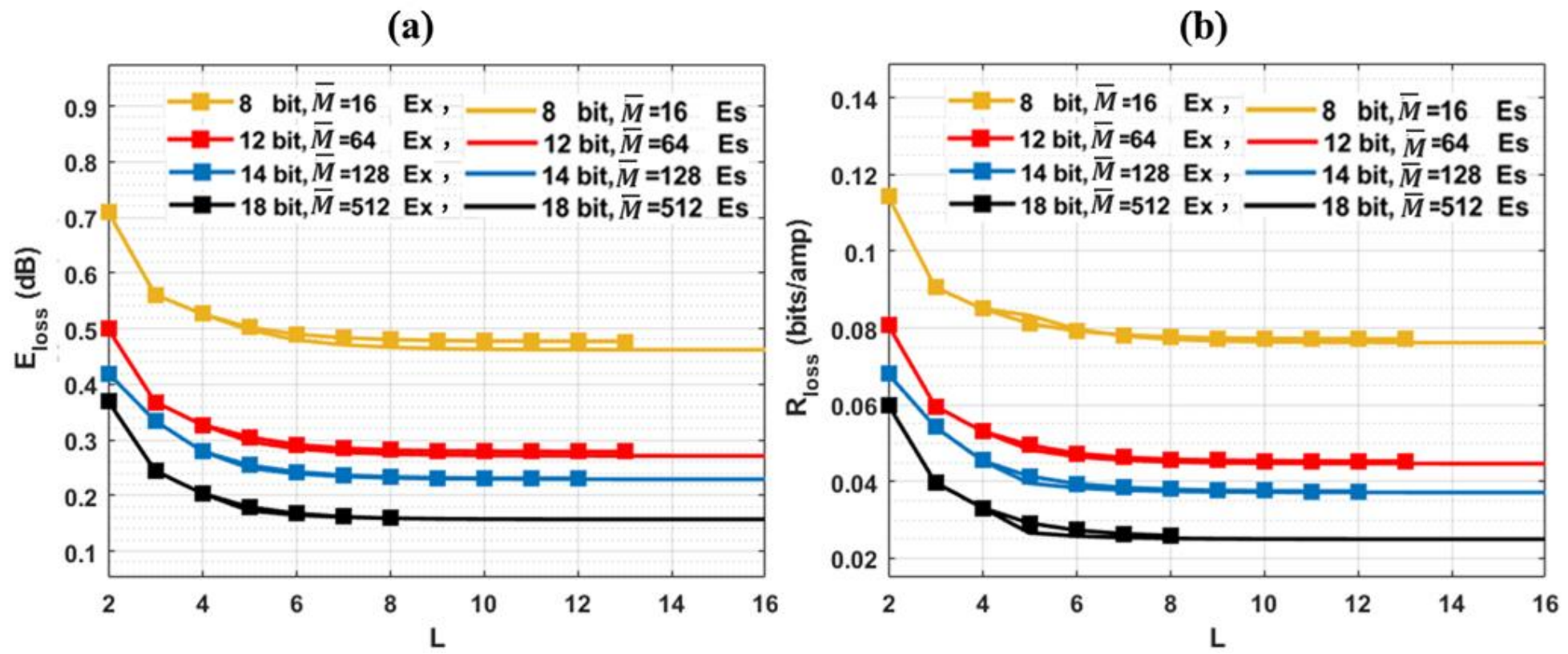

**Fig. 3.** (a) Energy loss and (b) rate loss as a function of the number of layers for $R_{DM} = 0.5$ (bits/amp) and $n = 2$. Straight lines represent equations (8) and (9), whether squares have been obtained through exhaustive search. Model estimation is shown for $N_b = 8{,}12{,}14{,}18$, corresponding to $\bar{M} = 16{,}64{,}128{,}512$. Es =Estimated, Ex = Exhaustive.

However, improvements slow significantly beyond a certain $L$.

For $L \geq 5$ and finite $N_b$ we propose the following general expression for the mean energy per amplitude of an $L$-layer structure:

$$E_{DM}^{(L)} = E_{DM}^{(4)} + \frac{E_{DM}^{(2)}}{4N^{(3)}}(2^{5-L} - 1) - \alpha \qquad (7)$$

where $E_{DM}^{(4)}$ and $E_{DM}^{(2)}$ are the mean energy per amplitude values for 4- and 2-layer HiDM structure, respectively, $N^{(3)}$ is the DM word of a 3-layer HiDM structure (all obtained through exhaustive search), and $\alpha = 0.014$ when $0.35 < R_{DM} \leq 70$. $E_{DM}^{(2)}$, $N^{(3)}$, and $E_{DM}^{(4)}$ are obtained for 2,3, and 4-layer optimal structures obtained after exhaustive search. Using eq. (6):

$$E_{loss}^{(L)} = 10\log_{10}\left(\frac{E_{DM}^{(L)}}{E_{MB}}\right) \qquad (8)$$

Similarly, $R_{loss}$ for an $L$-layer structure can be expressed as:

$$R_{loss}^{(L)} = R_{loss}^{(4)} + \frac{R_{loss}^{(2)}}{4N^{(3)}}(2^{8-L} - 1) - 0.008 \qquad (9)$$

where $R_{loss}^{(4)}$ and $R_{loss}^{(2)}$ are the rate loss values for 4- and 2-layer HiDM structure, respectively, and $N^{(3)}$ is the DM word of a 3-layer HiDM structure (all obtained through exhaustive search). Equations (8) and (9) are validated through exhaustive search, and the results are shown in Fig.3, where $E_{loss}$ and $R_{loss}$ are calculated for $R_{DM} = 0.5$ (bits/amp) and $n = 2$. Model estimation is shown for $N_b = 8{,}12{,}14{,}18$ corresponding to $\bar{M} = 16{,}64{,}128{,}512$. Straight lines represent equations (8) and (9), whether squares have been obtained through exhaustive search, and are limited to the specific number of layers $L$ for each configuration due to computational memory resources. The agreement between model and exhaustive search is quite good, with an error below 3.3% and 1.8% in the estimation of $E_{loss}$ and $R_{loss}$ respectively in the worst case of 8 bits. A small number of $N_b$ ($N_b < 8$) leads to a higher $E_{loss}$ and $R_{loss}$. Moreover, from equations (8) and (9), $\lim_{L\to\infty} E_{loss}^{(L)} > 0$ and $\lim_{L\to\infty} R_{loss}^{(L)} > 0$ providing lower bounds as shown in the graphs. Finally, it can be noticed that for all considered cases, an increase in the number of layers above 8 does not give any significant loss reduction.

Using our estimation method, for a 19-layer structure with $R_{DM} = 0.5$ (bits/amp) and 12 bits, we obtained $R_{loss} = 0.044$ (bits/amp). However, achieving this requires a block length of N=2,883,584, which is impractically large, and the resulting rate loss remains non-negligible. By tuning the finesse coefficient $\alpha$ in (7), the model can be made to accurately match the data at both very low and very high distribution matcher (DM) rates. To obtain an accurate lower bound on the energy loss $E_{\text{loss}}$ across these regimes, $\alpha$is defined as follows:

$$\begin{cases} R_{DM} \leq 0.35, & \alpha = 0.007 \\ 0.70 < R_{DM}, & \alpha = 0.028 \end{cases}$$

To validate the accuracy of the proposed model across different rates, its performance was evaluated over a range of $R_{\text{DM}}$values. Using (8) and (9), lower bounds on the energy loss $E_{\text{loss}}$and rate loss $R_{\text{loss}}$were derived for HiDM structures with $R_{\text{DM}} = 0.25$, 0.625, and 0.75, all with $N_b = 12(\bar{M} = 64)$. The corresponding results are shown by the dashed-dotted curves in Figs. 4(a) and 4(b), respectively. As illustrated in Fig. 4, the proposed model provides lower-bound estimates of $E_{\text{loss}}$and $R_{\text{loss}}$that closely match the results obtained via exhaustive evaluation.

### D. Memory Estimation

The total memory needed to store all the LUTs directly measures the structure's complexity in the context of HiDM. Considering a full parallelization of each layer, the total required memory (in bits) covers the LUTs at the encoding side ($mem_{enc}$) and decoding side ($mem_{dec}$), as proposed in [10]. Consequently, the design of HiDM must address the trade-off between performance and required memory. Since decoding memory is always higher than encoding memory, as a worst-case scenario $2mem_{dec}$ can be considered as a good estimation of the requested hardware. Considering the number of LUTs in each layer, the decoding memory can be obtained by the following equation:

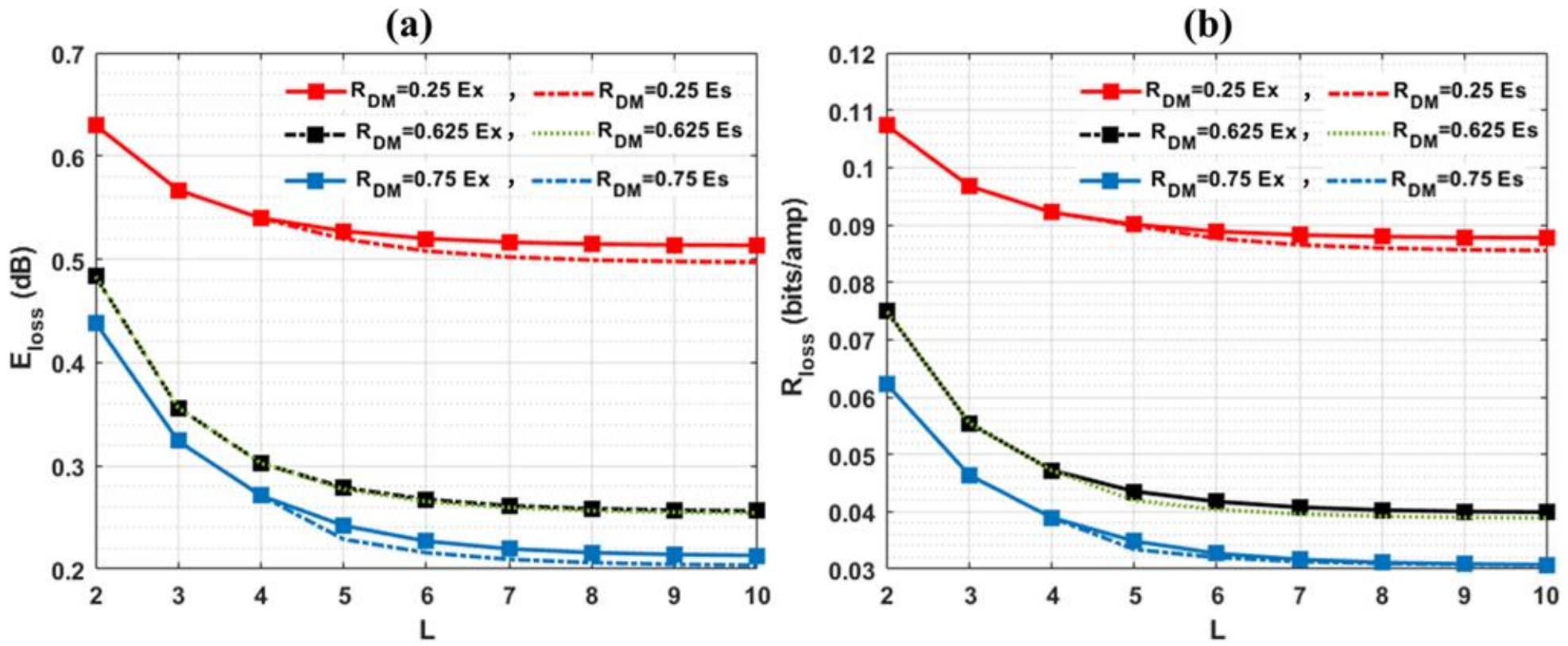

**Fig. 4.** Obtained lower bound by exhaustive search (marked by squares) and obtained lower bound based on the proposed design rule (dashed-dot lines) for 12-bit structures with maximum possible $\bar{M}$ and with RDM, of (a) 0.25, (b) 0.625, and 0.75. Tuning $\alpha$, the results are perfectly matched with an exhaustive one.

$$mem_{dec} = T_1\, 2^{N_1}(\vartheta + k_1) + \sum_{l=2}^{L-1} T_l\, \bar{M}^2(\vartheta + k_l) + \bar{M}^2 k_L \quad (10)$$

Based on our model, $T_l$will be simplified to $T_l = n^i$ , where $i = L - l$ , which indicates the number of times that the $l^{th}$ layer in the structure will be used, results of full parallelization (with $T_L = 1$).

However, since our design rules don't provide us all $k$ values for $l > 4$ in advance, we can only calculate the first terms of the equation. Note that the distribution of $\boldsymbol{k}$ varies with $L$, in which the $\boldsymbol{k}$ of a 4-layer structure will be different from that of the first 4-layer of a structure with 7 layers. For this reason, we assume $k_{l>4} = 1$, which simplifies the computation and enables a partial estimation of the decoding memory. It should be emphasized that, due to the construction of the HiDM architecture, the condition $k_{l>4} = 1$ is not feasible for all structures. This assumption is therefore adopted solely to facilitate a more efficient estimation of the decoding memory. Fig. 5(a) illustrates the decoding memory (in bits) as a function of the number of layers for $R_{\text{DM}} = 0.5$bits/amp and $n = 2$, considering different values of $N_b$. Fig. 5(b) shows the decoding memory variation of the optimized 7-layer structures as a function of the number of bits. In both cases, the estimated results closely match the actual memory requirements.

*E. Proposed algorithm for system design*

Algorithm 1 outlines the process for obtaining the optimal HiDM.

**Algorithm1.** Design HiDM

1. *Define DM rate*
2. *Set $N_b$ to the maximum of SYL*

*Obtain Eq 7 (or 9) parameters ($E_{loss}^{(4)}$, $N^{(3)}$ , $E_{loss}^{(2)}$,(or $R_{loss}^{(4)}$, $N^{(3)}$ , $R_{loss}^{(2)}$ ) in Steps 3 and 4:*

3. *Optimize characterization vectors*
   - *a)* $L = 4$
   - *b)* $\vartheta = \left\lceil \frac{N_b}{n} \right\rceil$, *and* $\boldsymbol{M} = (2, \bar{M}, \bar{M}, \bar{M})$ , *(where* $\bar{M} = 2^{\vartheta}$ *)*
   - *c) n=2*
   - *d)* $\boldsymbol{N} = (N_1, n, n, n)$ , *the DM word is equal to* $N_1 n^{L-1}$
   - *e) given $R_{DM}$, run an exhaustive search to find $\boldsymbol{k}$ vector with:*
     - *fixed $\boldsymbol{M}$ and N= $((N_b - c), n, n, n)$:*
       - *i. Start search with $c = 0$,*
       - *ii. Vary c from 0 to $N_b - 1$ with unit step*
       - *iii. Stop searching for c when $R_{loss}$ at $N_1 = N_b - (c - 1)$ exceeds that at $N_1 = N_b - c$, with $c \in \mathbb{N}$. Use $N_1 = N_b - c$ to determine the optimal $\boldsymbol{k}$.*
4. *Repeat step 3 with L=3 and L=2 to generate 2 and 3-layer structures*

*Estimate L for the target $E_{loss}$ (or $R_{loss}$):*

5. *Use equations 7 (or 9) to estimate the lower bound of $E_{loss}$ (or $R_{loss}$) as a function of the number of layers*
6. *From the output of step 5, select the number of layers $L_s$ just before $E_{loss}$ (or $R_{loss}$) saturation*

*Verify memory:*

7. *Estimate memory (using Eq 10)*
   - *i. If the estimated memory suits the limitation, run the exhaustive search to find the $\boldsymbol{k}$ vector for the L-layer structure, having $\boldsymbol{N}$ and $\boldsymbol{M}$ from step 3*
   - *ii. If the estimated memory is beyond the limitation, either reduce the number of layers or repeat Steps 3 to 7 with $\bar{M} = 2^{\vartheta-1}$, if memory suits the limitation, find the $\boldsymbol{k}$ vector through an exhaustive search; otherwise, repeat Step 7.*

---

To provide a more practical evaluation of the performance of our designed HiDM, we consider a 12-bit synthesizer limitation (SYL) that is compatible with typical hardware resources and an 8 MB memory constraint. We generated 2, 3, and 4-layer structures with $R_{DM} = 0.75$ (bits/amp), $N_b = 12, n = 2, \bar{M} = 64$, and obtained $E_{loss}^{(4)}$, $N^{(3)}$ , $E_{loss}^{(2)}$, 0.438 dB, 44, and 0.027 dB, respectively. Using this information, we derive a lower bound on $E_{loss}$ (Eqs. 7-8 and $\alpha = 0.028$) and $R_{loss}$ (Eq. 9) as a function of the number of layers (Fig. 6).

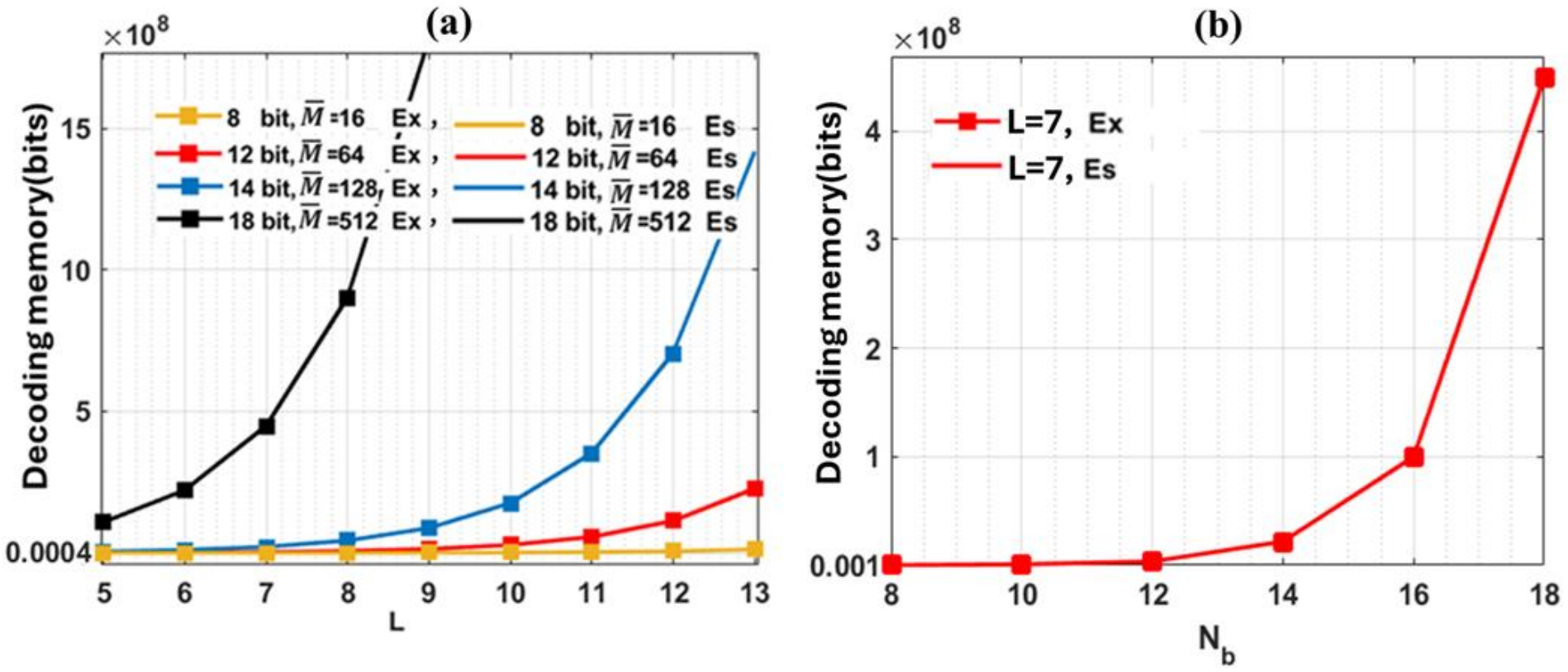

**Fig. 5.** Decoding memory (bits) (a) as a function of the number of layers, (b) as a function of the number of bits, for $R_{DM} = 0.5$ (bits/amp) and $n = 2$. Straight lines represent equation (10), assuming $k_{l>4} = 1$ , whether squares have been obtained through exhaustive search. Model estimation is shown for $N_b = 8,10,12,14,16,18$, corresponding to $\bar{M} = 16,32,64,128,256,512$. Es =Estimated, Ex = Exhaustive.

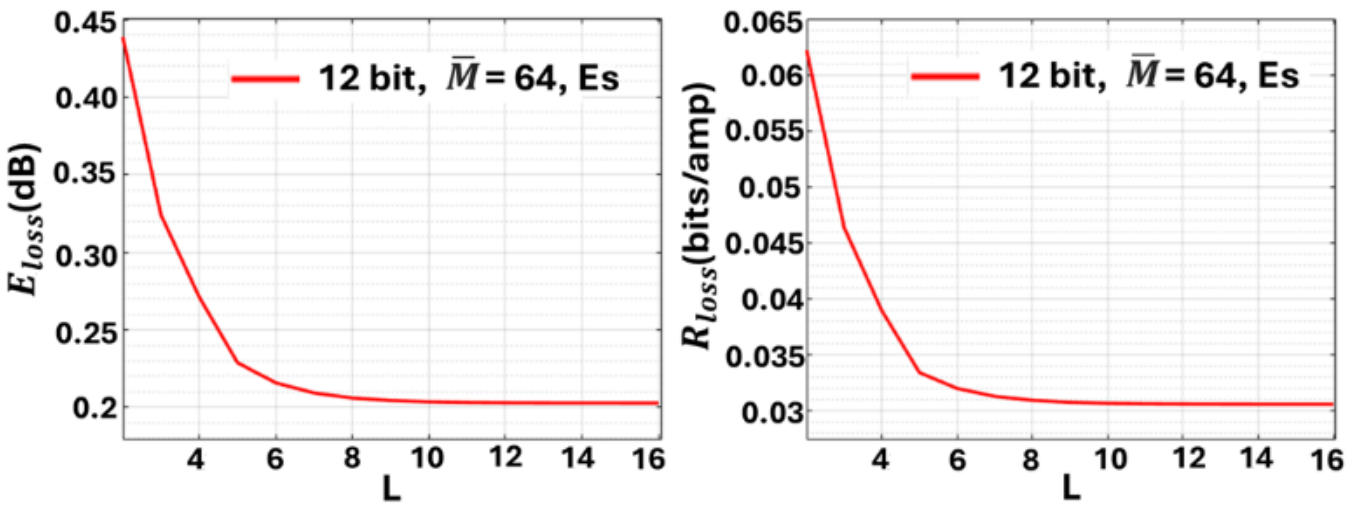

**Fig. 6.** Energy(a) and Rate loss(b) obtained according to Algorithm 1.

Based on the lower-bound analysis of $E_{\text{loss}}$, a 7-layer design is expected to achieve $E_{\text{loss}} = 0.2093$dB and $R_{\text{loss}} = 0.0313$bits/amp. Introducing an eighth layer yields only marginal improvements, reducing $E_{\text{loss}}$by approximately $4 \times 10^{-3}$dB and $R_{\text{loss}}$by approximately $4 \times 10^{-4}$bits/amp. Therefore, a 7-layer architecture ($L_s = 7$) was selected, and the corresponding optimal structure was determined using Step 3 of Algorithm 1. Prior to the final exhaustive search over $N_1$and $\boldsymbol{k}$, the decoding memory requirement was estimated to be approximately 7.9 MB (2 $\text{mem}_{\text{dec}}$), which is well within the imposed constraints.

Following the exhaustive optimization, the final structure parameters were obtained, resulting in $E_{\text{loss}} = 0.21$dB and $R_{\text{loss}} = 0.0316$bits/amp. These results are in good agreement with the analytical estimates. The estimated decoding memory exhibits only a 0.72% relative error with respect to the optimized 7-layer structure, with $mem_{dec,est} = 3\,977\,216 bits$ and $mem_{dec,act} = 3\,948\,544 bits$.

The slight overestimation arises from differences in the $k$-parameter configuration between the reference structure used for estimation and the final optimized design. Specifically, the 4-layer reference structure employs $k_{\text{4-layer}} = (4,5,3,8)$, whereas for the 7-layer configuration, we initially assumed $k = 1$for layers $l = 5,6,7$. The actual optimized 7-layer structure uses $k_{\text{7-layer}} = (4,4,5,4,4,4,8)$. Since the optimized 7-layer design exhibits smaller $k$ values in the intermediate layers than those of the 4-layer reference structure, its actual decoding memory requirement is lower, thereby explaining the observed overestimation in the analytical model.

## III. Performance Evaluation

We evaluated the performance of the proposed HiDM structure that incorporates the selected layer based on the lower-bound $E_{loss}$ information, for the PAS implementation of an AWGN channel, in terms of the OSNR dependence of normalized generalized mutual information (NGMI) in the linear regime, compared to [21]. Notably, $H$ (bits/amp) refers to the entropy of shaped amplitudes (1D Symbol); thus, the information rate of the probabilistic shaped constellation, per polarization, can be given by [22]:

$$R_{PAS_QAM} = H(P) - m(1 - R_{FEC}) \qquad (11)$$

where $H(P) = (2H + 2)$ represents the entropy of constellation distribution (bits/QAM symbol), $R_{FEC}$ refers to a given forward-error-correction (FEC) rate and $m = \log_2(16)$ for 16QAM. To investigate the performance evaluation of our proposed method, we compared our design performance with the previously reported PAS-16QAM constellation [21] with a rate of 0.75 bits/amp (constellation rate 3.5 bits/QAM symbol) and 25% FEC overhead for a 200 Gb/s net data rate, where the target information rate will be 2.70 bits/QAM symbol. We applied our method to generate a HiDM structure for a constellation that matches the transmission rate of [21]. Fig 7 shows NGMI versus OSNR, enabling a direct performance comparison between uniform and probabilistically shaped 16QAM operated at different baud rates, as in [21], and our proposed structure. To ensure a fair comparison at a net target rate of 200 Gb/s, the uniform and probabilistically shaped 16-QAM modes are evaluated at different symbol rates. An OSNR gain of 1.542 dB is achieved for the shaped constellation at a symbol rate of 37 Gbaud, with a rate loss of $R_{\text{loss}} = 0.0316$ bits/amp and an energy loss of $E_{\text{loss}} \approx 0.21\text{dB}$. Although the shaping method in [21] is not explicitly specified, the proposed 7-layer HiDM structure outperforms the reported results by 2.8% in terms of shaping gain.

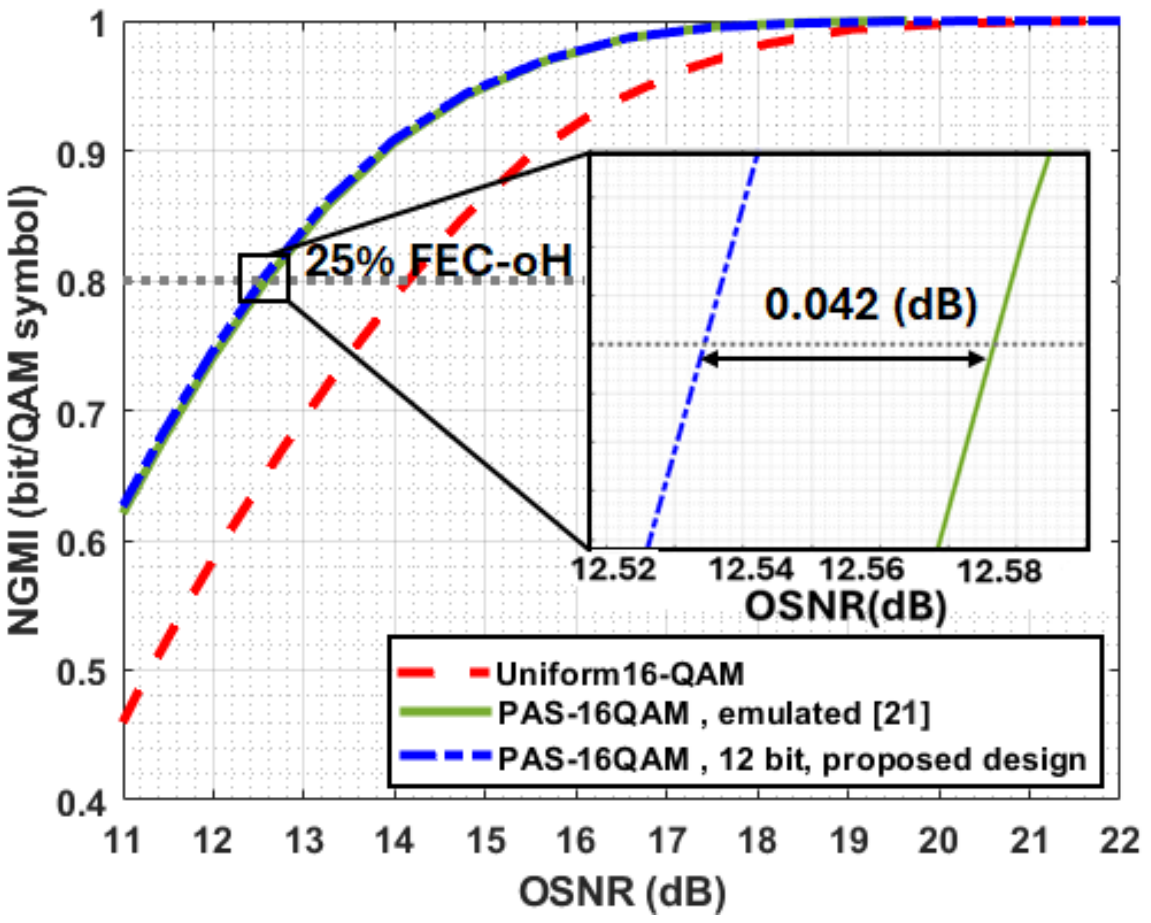

**Fig. 7.** NGMI as a function of OSNR. PAS-16QAM performance of 12-bit HiDM using our design rule (blue) and the emulated result from [21] (green). The zoomed-in view shows a 0.042 dB gain, as per our design rule.

## V. Conclusion

This paper presented a semi-analytical design methodology for hierarchical distribution matchers (HiDMs) targeting the Maxwell–Boltzmann (MB) distribution. The proposed approach enables the optimization of key HiDM parameters, including the number of input bits $k$, the number of output amplitudes $N$, and the alphabet size $M$, thereby reducing shaping and energy gaps and allowing operation closer to channel capacity. Unlike conventional HiDM design methods, which rely on exhaustive parameter searches that become computationally prohibitive as the number of layers increases, the proposed framework significantly reduces both search complexity and hardware resource requirements. In particular, a 98.56% reduction in the number of parameter combinations was demonstrated for the optimization of a 4-layer structure with 12 input bits, reducing the search space from 20,417 to 294 configurations.

In conventional HiDM architectures, rate loss and energy loss are evaluated after constructing the full structure, based on the entropy and probability distribution of the output amplitudes. In contrast, the proposed model enables the estimation of lower bounds on rate loss and energy loss prior to full HiDM construction, as well as the identification of the minimum number of layers required to avoid unnecessary architectural complexity. It was shown that, under practical hardware constraints, the HiDM rate loss remains strictly above its lower bound and does not vanish; equivalently, the mean symbol energy remains above its theoretical minimum.

Furthermore, the proposed framework allows for the estimation of decoding memory requirements, which dominate over encoding memory. In the worst-case scenario, a decoding memory on the order of 2 $\text{mem}_{\text{dec}}$ provides a reliable estimate of the required hardware resources. This capability facilitates the design of HiDM structures that comply with contemporary hardware and design-tool constraints, including those associated with modern application-specific integrated circuits (ASICs) and field-programmable gate arrays (FPGAs). The accuracy of the proposed model was validated for probabilistic amplitude shaping of 16-QAM (PAS-16QAM) across a wide range of bit rates and distribution matcher rates, showing close agreement between analytical predictions and actual values. Finally, the performance of the optimized HiDM structure was evaluated over an AWGN channel in terms of normalized generalized mutual information (NGMI) as a function of the optical signal-to-noise ratio (OSNR). A shaping gain improvement of 2.8% was demonstrated at a net data rate of 200 Gb/s with 25% forward error correction (FEC) overhead, compared to previously reported results.

**Disclosures.** The authors declare no conflicts of interest.